\documentclass[prb,preprint, superscriptaddress]{revtex4-1}

\usepackage[dvipdfmx]{graphicx}
\usepackage{dcolumn}% Align table columns on decimal point
\usepackage{bm}% bold math

\usepackage{color}

\begin{document}

% \preprint{APS/123-QED}

\title{One-dimensionality of the spin-polarized surface conduction and valence bands of quasi-one-dimensional Bi chains on GaSb(110)-(2$\times$1)}

\author{Yoshiyuki Ohtsubo}
\email{y\_oh@fbs.osaka-u.ac.jp}
\affiliation{Graduate School of Frontier Biosciences, Osaka University, Suita 565-0871, Japan}
\affiliation{Department of Physics, Graduate School of Science, Osaka University, Toyonaka 560-0043, Japan}
\author{Naoki Tokumasu}
\affiliation{Department of Physics, Graduate School of Science, Osaka University, Toyonaka 560-0043, Japan}
\author{Hiroshi Watanabe}
\affiliation{Graduate School of Frontier Biosciences, Osaka University, Suita 565-0871, Japan}
\affiliation{Department of Physics, Graduate School of Science, Osaka University, Toyonaka 560-0043, Japan}
\author{Takuto Nakamura}
\affiliation{Department of Physics, Graduate School of Science, Osaka University, Toyonaka 560-0043, Japan}
\author{Patrick Le F\`evre}
\author{Fran\c{c}ois Bertran}
\affiliation{Synchrotron SOLEIL, Saint-Aubin-BP 48, F-91192 Gif sur Yvette, France}
\author{Masaki Imamura}
\author{Isamu Yamamoto}
\author{Junpei Azuma}
\author{Kazutoshi Takahashi}
\affiliation{Synchrotron Light Application Center, Saga University, Yayoigaoka 8-7, Tosu, Saga 841-0005, Japan}
\author{Shin-ichi Kimura}
\email{kimura@fbs.osaka-u.ac.jp}
\affiliation{Graduate School of Frontier Biosciences, Osaka University, Suita 565-0871, Japan}
\affiliation{Department of Physics, Graduate School of Science, Osaka University, Toyonaka 560-0043, Japan}

\date{\today}

\begin{abstract}
Surface electronic structure and its one-dimensionality above and below the Fermi level ($E_{\rm F}$) were surveyed on the Bi/GaSb(110)-(2$\times$1) surface hosting quasi-one-dimensional (Q1D) Bi chains, using conventional (one-photon) and two-photon angle-resolved photoelectron spectroscopy (ARPES) and theoretical calculations.
ARPES results reveal that the Q1D electronic states are within the projected bulk bandgap.
Circular dichroism of two-photon ARPES and density-functional-theory calculation indicate clear spin and orbital polarization of the surface states consistent with the giant sizes of Rashba-type SOI, derived from the strong contribution of heavy Bi atoms. 
The surface conduction band above $E_{\rm F}$ forms a nearly straight constant-energy contour, suggesting its suitability for application in further studies of one-dimensional electronic systems with strong SOI.
A tight-binding model calculation based on the obtained surface electronic structure successfully reproduces the surface band dispersions and predicts possible one- to two-dimensional crossover in the temperature range of 60--100~K.
\end{abstract}

\maketitle

\section{Introduction}

One-dimensional (1D) and quasi-1D (Q1D) systems have emerged as some of the most interesting candidates for studying the non-conventional electronic phenomena caused by reduced degrees of freedom, such as various quantum fluids replacing the broken-down Fermi-liquid (FL) model \cite{Voit95, Imambekov09, Grioni09}.
Among these systems, spin-split ones such as the 1D edge states of two-dimensional topological insulators (TI) \cite{Hasan10, Konig07} are attracting significant attention because of their possible applications in spintronic low-dimensional devices \cite{Pesin12}, as well as the expected exotic electronic phenomena such as the formation of Majorana bound states \cite{Mourik12} and the rich phase diagram based on Tomonaga--Luttinger liquid \cite{Voit95, Sedlmayr13}.
In recent days, intermediate materials changing from 1D non-FL quantum fluid to ordinary 3D FL depending on temperature have also been discovered, providing an attractive playground to study the electronic behavior in dimensional-crossover systems \cite{Biermann01, Nicholson17}.

The Bi/GaSb(110)-(2$\times$1) surface is one of the promising candidates to study such 1D electronic phenomena with significant contributions from the spin-orbit interaction (SOI), because the Bi-(2$\times$1) chains on the (110) surfaces of III--V semiconductors \cite{Ford90} host the Q1D surface valence bands (VBs) with giant sizes of spin splitting derived from Rashba-type SOI \cite{Nakamura18, Nakamura19, Rashba}.
However, the observed band dispersions of the Bi-(2$\times$1) surface are limited below the Fermi level ($E_{\rm F}$), although the conduction bands (CBs) above $E_{\rm F}$ are also important, such as for practical applications of spin-dependent transport phenomena.
Moreover, the obtained spin-split VBs form an almost linear Fermi contour (FC), indicating a small, but finite contribution from a two- or three-dimensional component, which would be an inter-chain coupling between surface Bi chains \cite{Nakamura18}.

In this paper, we report the surface electronic structure not only below but also above the $E_{\rm F}$ on a Bi/GaSb(110)-(2$\times$1) surface using conventional (one-photon) angle-resolved photoelectron spectroscopy (ARPES) with synchrotron radiation and two-photon ARPES with a pulse laser, together with theoretical calculations.
ARPES results show the Q1D electronic states both below and above $E_{\rm F}$ within the projected bulk bandgap.
Circular dichroism (CD) of two-photon ARPES and density-functional-theory (DFT) calculations indicate clear spin and orbital polarization of the surface states, as expected from the Rashba-type SOI.
The surface CB above $E_{\rm F}$ forms a nearly straight constant-energy contour, suggesting it is suitable for application in further studies of one-dimensional electronic systems with strong SOI.
We also examine possible 1D to 2D crossover in surface electronic states of Bi/GaSb(110)-(2$\times$1) based on the obtained surface electronic structure, suggesting the possible emergence of 1D to 2D crossover in the range of 60--100 K on the surface CBs.

\section{Methods}

The GaSb(110)-(1$\times$1) substrates (nominally undoped) were cleaned using repeated cycles of Ar ion sputtering (1 keV) and annealing at $\sim$700 K, resulting in a sharp (1$\times$1), low-energy electron diffraction (LEED) pattern, as shown in Fig. 1 (a).
A few monolayers of Bi were then evaporated from a homemade Knudsen cell at room temperature, and subsequent annealing up to 550 K formed the (2$\times$1) surface reconstruction as shown in Fig. 1 (b).
The rectangular surface unit cell with anistropic lengths of the surface unit vectors is consistent with various surface 1D systems \cite{Rowe75, Abukawa95}.
This procedure is similar to a preparation method used in the earlier work \cite{Nakamura19}.

ARPES measurements were performed with synchrotron radiation at the CASSIOP\'EE beamline of the synchrotron SOLEIL ($h\nu$ = 30 eV, circularly polarized).
Two-photon ARPES experiments were conducted at the Saga University beamline BL13 at the Saga Light Source.
We used laser pulses with $h\nu$ = 1.5 eV for pump and 6.0 eV for probe pulses, respectively, generated from a Ti:sapphire laser source (100 kHz, 340 mW, 1.5 eV, 200 fs).
The power of the incident pump pulses was set to 2.0 mW monitored by a wide-band thermopile sensor.
For the probe pulses, its power was monitored from the photoelectron total yields from the Bi/GaSb(110) samples (10$\pm$2 pA), since its intensity is too low to be detected by the same manner as the pump pulses (at most, it is lower than 1 $\mu$W).
The pump (probe) laser pulses were linearly (circularly) polarized. %, sharing a common photon-incident plane to the sample.
The origin of the time delay (time-zero) was determined based on the cross-correlation profile between the pump and probe laser pulses by detecting photoelectrons; the overall time resolution of the current two-photon ARPES setup is estimated to be 0.4 ps from the same data.
The energy resolutions of the conventional and two-photon ARPES in this work were $\sim$15 meV and $\sim$35 meV, respectively.
The energy resolution and photoelectron kinetic energies at $E_{\rm F}$ were calibrated using the Fermi edge of the photoelectron spectra from polycrystalline gold electrically attached to the samples.
In the two-photon ARPES experiments, we applied a bias voltage of -15 V to the sample to expand the field of view in momentum space.
The distortion in the ARPES image is corrected according to ref. \onlinecite{Hengsberger08}.
This correction method was confirmed by comparing the known surface bands (the cleaved Bi$_2$Te$_3$(0001) surface \cite{Bi2Te3}) observed with and without the sample bias.
It was also checked that there was no  multi-photon excitation by pump pulses and that the space-charge effect by the probe photons were negligibly small in the two-photon ARPES measurements.

Figure 1 (c) shows the experimental geometry of the one-photon and two-photon ARPES measurements in this work.
Small circles represent the Bi chain orientation.
The photon incident plane is ($\bar{1}$10) and it is normal to the photoelectron detection plane.
In the two-photon ARPES experiment, both pump and probe photons shared the incident plane ($\bar{1}$10) for two-photon ARPES.
The electric field of the linearly polarized pump pulse lied in the photon-incident plane and the definition of the left- and right-handed circular polarizations are depicted as L and R in Fig. 1 (c), respectively.
Photon incident angle $\theta_i$ was set from 40$^{\circ}$ to 70$^{\circ}$ by rotating the polar angle of the sample.
The deflector angle of the photoelectron analyzer is also used to obtain the constant-energy contour as Figs. 2 (a) and 3 (a).

DFT calculation was performed using WIEN2k code with SOI taken into account\cite{W2k}.
The surface was modeled by a symmetric slab of 20$\times$2 GaSb layers, with the surface covered with (2$\times$1) zigzag Bi chains, and the atomic positions of four Bi atoms and GaSb slabs down to three atomic layers from the topmost surface were relaxed using generalized-gradient approximations \cite{GGA_PBE}. The optimized structure agreed well with that determined previously \cite{Gemmeren98}.
To calculate the surface electronic structure, we used the modified Becke--Johnson potential together with the exchange-correlation potential constructed using the local density approximation \cite{mBJ1, mBJ2, Camargo12}.

\section{Results and Discussion}

\subsection{ARPES band dispersions}

Figure 2 shows the VB dispersions of Bi/GaSb(110)-(2$\times$1) by ARPES with circularly polarized photons ($h\nu$ = 30 eV).
Along both $\bar{\Gamma}$-$\bar{\rm X}$ (Fig. 2 (b)) and $\bar{\rm Y}$-$\bar{\rm M}$ (Fig. 2 (c)), paired upward-convex parabolic bands were observed.
These bands are absent on the clean GaSb(110) surface \cite{Manzke87}; thus, they are obtained from the Bi-(2$\times$1) surface superstructure.
These dispersions are similar to each other and form a nearly 1D Fermi contour, as shown in Fig. 2 (a), indicating the surface's Q1D nature.
However, there are finite, but non-negligible, dispersions perpendicular to the surface chains, which make the top of the surface VB slightly below $E_{\rm F}$ around $\bar{\Gamma}$ but above it around $\bar{\rm Y}$.
%The 2D contribution is also suggested from the waving shape of the Fermi contour itself.
These features are consistent with the earlier work reporting the surface VBs on Bi/GaSb(110)-(2$\times$1) \cite{Nakamura18, Nakamura19}.

We also performed the two-photon ARPES, revealing the electronic structure above $E_{\rm F}$, as shown in Fig. 3.
The bottoms of the obtained CBs are located at $E_{\rm B}\sim$ 0.45 eV around both  $\bar{\Gamma}$ (Fig. 3 (b)) and $\bar{\rm Y}$ (Fig. 3 (c)).
The observed CBs lie in the projected bulk band gap \cite{Carstensen90}.
Together with the straight constant-energy contour shown in Fig. 3 (a), it is strongly suggested that this CB is also from the Bi-(2$\times$1) surface, not from the isotropic bulk electronic structure.
Moreover, the observed CB dispersions agree well with the paired parabolic ones, as guided by the curves in Figs. 3 (b) and 3 (c).
Such behavior is expected for the spin-split surface states due to the Rashba-type SOI, and is discussed in the following part (Sec. III B).

\subsection{Circular dichroism of ARPES}

To obtain experimental insight into the possible contribution from Rashba-type SOI to the CBs observed above, we analyzed the CD of the two-photon ARPES intensities as summarized in Fig. 4, based on the same dataset as those shown in Figs. 3 (b) and 3 (c).
Based on the ARPES geometry (Fig. 1 (c)), the CD signals should represent the orbital-angular-momentum (OAM) polarization of the photo-excited states perpendicular to [1$\bar{1}$0], which is of [001] or [110] orientation \cite{Wang11}.
It should be noted that the ``photo-excited state'' in this case is the electronic state before receiving the probe pulses, but after the pump ones.
Assuming the in-plane potential gradient, OAM, and spin polarizations along both orientations, the in-plane [001] and out-of-plane [110] are consistent with the spin-orbital splitting derived from Rashba-type SOI \cite{Premper07}.

Figure 4 (a) shows the CD signal $P$ = $(I_R - I_L)/(I_R + I_L)$ along $\bar{\Gamma}$-$\bar{\rm X}$ at each $E_{\rm B}$.
$I_R$ ($I_L$) is proportional to the photoelectron intensities obtained with right- (left-) handed probe photon polarizations.
$I_R$ and $I_L$ are obtained by dividing each raw spectra by the integrated two-photon ARPES intensity in the 2D region shown in Fig. 3 (b) so that the total photoelectron yield influenced by the intensity variations of probe photons are canceled out.
From the obtained polarizations in Fig. 4 (a), one can find that there is a positive peak on the left side and that there are common, linear backgrounds proportional to $k_{//[1\bar{1}0]}$, independent of $E_{\rm B}$.
To remove the contribution from these backgrounds, we derived the background shape from the fitting of the $P$ spectrum at $E_{\rm B}$ = 0.2 eV, where there are obviously no bands, and obtained the dashed line overlaid in Fig. 4 (a).
This background is consistent in the whole of the CB region shown in Fig. 4.
Although its origin is not clear, the surroundings of the sample (polycrystalline metal parts, such as Mo and Au) might provide such uniformly polarized photoelectrons.
After the background subtraction, we obtained the series of $P$ spectra as shown in Fig. 4 (b), where small but non-zero negative signals appear on the right sides ($k_{//[1\bar{1}0]} > 0$).
This trend is also visible around $\bar{\rm Y}$, as shown in Fig. 4 (c), obtained by background subtraction in the same manner as Fig. 4 (b).
Figures 4 (d) and 4 (e) are the CD plots based on the same analysis as those for Figs. 4 (b) and 4 (c), respectively.
These are based on $P$ and photoelectron intensities as illustrated in Fig. 4 (f), where the paired parabolic dispersion and its OAM polarizations are also shown.
The signs of the OAM polarizations invert with respect to the center of SBZ.
This characteristic agrees well with what is expected for the Rashba-type SOI.

% alpha-R for conduction bands
The size of the Rashba-type SOI is often characterized by the Rashba parameter $\alpha_{\rm R}$, which is the proportional constant with $k$ to determine the size of the spin-orbital splitting of the surface bands.
$\alpha_{\rm R}$ can be obtained from the energy and $k$ differences from the Kramers point.
Based on the observed dispersion of the surface bands of Bi/GaSb(110)-(2$\times$1) (Figs. 2 and 3), $\alpha_{\rm R}$ for the surface VBs are evaluated as 4.1 (2.6) eV\AA \ around $\bar{\Gamma}$ ($\bar{\rm Y}$), which is consistent with the earlier result \cite{Nakamura19}.
In contrast to the different $\alpha_{\rm R}$ around the different high-symmetry points in VBs, a common value of $\alpha_{\rm R}$ = 4.7 eV\AA \ could reproduce the observed CB dispersions around both $\bar{\Gamma}$ and $\bar{\rm Y}$.
The common $\alpha_{\rm R}$ value in SBZ suggests the smaller size of the 2D-term contribution to the surface CBs than that to the surface VBs, where $\alpha_{\rm R}$ is about 1.6 times different.
Although the CBs away from $\bar{\Gamma}$ are no longer parabolic, as shown by the dashed line in Fig. 3 (b), it is a common case for various electronic states that the dispersion deviates away from a simple parabolic one around high-symmetry points in a Brillouin zone \cite{Kittel}.

It is known that pump pulse sometimes cause the photo-induced modification in the electronic structure of materials \cite{PhotoPhase}.
However, although we swept the delay time between the pump and probe laser pulses, no clear difference in the CB dispersion was observed around either $\bar{\Gamma}$ or $\bar{\rm Y}$.
We have also observed that the CD has no clear delay-time dependence around $\bar{\Gamma}$.
These results suggest that the observed CBs and their CD reflect the dispersion and polarization of the unoccupied state, and that the photoexcitation and relaxation processes on the timescale of $\sim$1 ps have no major effect on these electronic states except for their electron populations.

\subsection{Surface bands by DFT calculation}

Figure 5 shows the calculated surface electronic structure of the Bi/GaSb(110)-(2$\times$1) surface.
Paired parabolic dispersions with strong contributions from surface Bi atoms were obtained, which were consistent with the ARPES experimental results shown above.
Based on the qualitative agreements of the observed and calculated band dispersions, and the clear contribution from the surface atoms in the calculated ones, it is confirmed that both the surface VBs and CBs observed by ARPES are derived from the surface Bi chains.
Quantitatively, the size of the bandgap between surface VBs and CBs is slightly overestimated, as shown on the bottom of the surface CB in Figs. 5 (a)--(d), due to the difficulty of estimating the bandgap of semiconductors via DFT \cite{Camargo12}, but it causes no problem in the following discussions.
Although the CBs are doubled around $\bar{\Gamma}$ as shown in Figs. 5 (a) and 5 (c), we found no significant difference between them in spin or OAM polarizations, or in atomic-orbital contributions.
Therefore, this doubling in the CBs is a kind of artifact from the imperfect symmetrization of the symmetric slab.

Figures 5 (a) and 5 (b) show that the surface VBs and CBs are clearly spin polarized with the sign inverting with respect to the center of SBZ, indicating that they are the spin-split bands due to the Rashba-type SOI.
Moreover, the OAM polarizations shown in Figs. 5 (c) and 5 (d) behave very similarly to the spin polarization; both surface CB and VB are mostly polarized to one orientation for one sign of $k_{//[1\bar{1}0]}$ and invert with respect to the center of SBZ.
The polarizations of the spin and OAM are anti-parallel in most region.
Such correspondence exhibits the strong coupling between spin and OAM due to SOI in the surface bands of Bi/GaSb(110)-(2$\times$1).
It is known that the CD of ARPES is sensitive to various experimental conditions, and thus one has to be careful that the observed signal is not always directly related to the spin polarization of the initial states \cite{Arrala13}.
It is true that the sign and amplitude of the CD signal could be modified due to the ARPES photoexcitation conditions, such as the photoelectron detection plane and probe photon energies.
However, in spite of such sensitivity of CD-ARPES signals, the calculated strong correspondence with the spin and OAM polarizations, together with the experimentally obtained CD signals and band dispersions, strongly supports the idea that the observed splitting and OAM polarization of the surface CBs are actually derived from the Rashba-type SOI.
We do not go into detail on the quantitative relationship between CD-ARPES signal and the OAM polarization of the initial state in this work because it is not necessary to confirm the spin-orbital splitting of the surface CB due to the Rashba-type SOI.

The dominant contribution from heavy Bi atoms on both the CBs and VBs also justifies the large size of Rashba-type SOI there.
One may find that the signs of the polarizations invert away from the center of SBZ in the calculated surface CBs.
These inversions are actually obtained in a recent theoretical calculation \cite{Nechaev19}.
It was not observed experimentally in this work (Fig. 4), possibly because it happens away from the center of SBZ at larger $E_{\rm B}$ regions that are out of the scope of the present two-photon ARPES measurements.

\subsection{One-dimensionality of the surface bands}

To examine the size of the 2D contribution to the obtained surface CBs and VBs, we used a simple tight-binding model,

\begin{equation}
E_k = -2t_a{\rm cos}(k_{//[1\bar{1}0]}a) -2t_b{\rm cos}(k_{//[001]}b) -\mu \pm\alpha_{\rm R}k_{//[1\bar{1}0]},
\end{equation}

where $t_a$ and $t_b$ are the effective hopping amplitudes along and perpendicular to the surface Bi chains, respectively, and $\mu$ is the chemical potential.
This model is the same as in the earlier work \cite{Nicholson17} except for the Rashba term $\alpha_{\rm R}k_{//[1\bar{1}0]}$, the sizes of which are fixed to the experimental values estimated in Sec. III. B.

Based on this model equation, we tried to reproduce the observed surface band dispersions and succeeded for both the surface VBs and CBs along $\bar{\Gamma}$-$\bar{\rm X}$, as shown by the dashed (solid) curves in Fig. 2 (3). 
The parameter sets for the tight-binding model ($t_a$, $t_b$, $\mu$) are summarized in Tab. I.
Moreover, it reproduced the dispersions along $\bar{\Gamma}$-$\bar{\rm Y}$ at the same time, as shown by the solid curves in Figs. 6 (a, b).
We also checked the model with the calculated bands based on DFT as shown in Fig. 6 (c).
The TB model could reproduce the obtained dispersions of surface VB and CB qualitatively, while an artificial splitting of the surface bands (discussed in the last section) and possible Rashba splitting along this orientation make the quantitative discussion in surface CB difficult.
Note that such splittings in surface CB are too small to be detected experimentally and that the TB model calculation could indicate the averaged dispersion of the split surface bands even with such splittings.
Based on the experimental data, the ratio between $t_a$ and $t_b$ for surface VB (CB) is $\sim$700 ($\sim$20), indicating significant 1D character of the surface bands; note that this ratio should be 1 for isotropic 2D or 3D bands.
The difference of the $t_a$/$t_b$ ratio for surface VB and CB also shows that the one dimensional character of surface CB is more significant than that of surface VB.
The general trend of $t_a$ and $t_b$ is identical for the DFT bands; $t_b$ is much smaller than $t_a$ indicating significant 1D character of surface CB.
Since the quantitative information based on DFT for semiconductors often lacks accuracy as discussed in sec. IIIC \cite{Camargo12}, we made the following discussion based on the prameter set for the experimental results.

The inter-chain bandwidth 4$t_b \equiv E_C$ is the characteristic energy scale at which the system transits from 1D to higher (2D) dimensional behavior, because the excitation with energies $E \gg E_C$ can ignore the 2D contribution, and thus, the whole system can be regarded as 1D \cite{Biermann01}.
The obtained model parameters $t_b$ are 55 and 5 meV for the surface VBs and CBs, respectively.
These energies correspond to the temperatures of 640 and 60 K, respectively, above which the electronic system would exhibit 1D, non-FL behavior.
It should be noted that the estimated energy scale $t_b$ = 5 meV for the surface CBs is smaller than the experimental energy resolution of the current two-photon ARPES.
Although 4$t_b$ = 20 meV is the expected bandwidth along [001] for the surface CBs, it could be as large as 35 meV, the energy resolution of the two-photon ARPES.
Assuming 4$t_b$ to be 20--35 meV, the dimensional-crossover temperature range is 60--100 K.
Unfortunately, it is not realistic to observe the dimensional crossover phenomena for the surface VBs across 640 K, because the Bi desorbs from the GaSb(110) surface at temperatures above $\sim$600 K.
For the surface CBs, neither the current method nor two-photon ARPES is suitable to observe the dimensional crossover, because the temperature of the electron system is always hot (typically above room temperature) and it is very difficult to distinguish the density of states around $E_{\rm F}$ of FL and non-FL cases at such a high temperature of the electron system; thermal broadening becomes dominant in both cases \cite{Nicholson17}.
To trace the possible dimensional crossover on Bi/GaSb(110)-(2$\times$1), it is desirable to fill the electrons to the surface CBs by the other static methods such as alkali-metal evaporation \cite{Hirahara09}.

At last, we examine the origin of the difference of the one dimensionalities among the surface bands.
The surface CB has much smaller influence from the inter-Bi-chain coupling than the surface VBs, as shown by the straight shape of the constant energy contour (Fig. 3 (a)) as well as  the inter-chain energy scale $t_b$ (5 meV for the CBs and 55 meV for the VBs).
To reveal the origin of this difference, we calculated the fractional orbital contributions of the surface Bi atoms at the surface VBs and CBs indicated by the arrows in Fig. 5 (a), as summarized in Table II.
The surface CBs and VBs have sizable contributions from both the Bi 6$s$ and 6$p$ orbitals.
Among these, the contribution from Bi-6$p_b$, the $p$ orbital with the lobe along [001], the inter-Bi-chain orientation, is significantly larger for the VBs.
This difference would be the origin of the higher influence of the 2D component in the surface VBs of Bi/GaSb(110)-(2$\times$1).

\section{Summary}

In this work, we report the surface electronic structure, both above and below $E_{\rm F}$, on a Bi/GaSb(110)-(2$\times$1) surface using conventional and two-photon ARPES and theoretical calculations.
ARPES results show the Q1D electronic states both below and above $E_{\rm F}$.
Circular dichroism of two-photon ARPES and DFT calculation indicate clear spin and orbital polarization of the surface CBs, as expected from the Rashba-type SOI.
The surface CB above $E_{\rm F}$ forms a nearly straight constant-energy contour, suggesting its suitability for application in further studies of one-dimensional electronic systems with strong SOI.
We also examined possible 1D to 2D crossover in surface electronic states of Bi/GaSb(110)-(2$\times$1) based on the obtained surface electronic structure, suggesting that the 1D to 2D crossover may appear at temperatures of 60--100 K.

\section{Acknowledgement}
We acknowledge F. Deschamps for her support during the experiments on the CASSIOP\'EE beamline at synchrotron SOLEIL.
This work was also supported by JSPS KAKENHI (grant numbers JP17K18757, JP19J21516,  and JP19H01830).

% LEED
\begin{figure}[p]
\includegraphics[width=80mm]{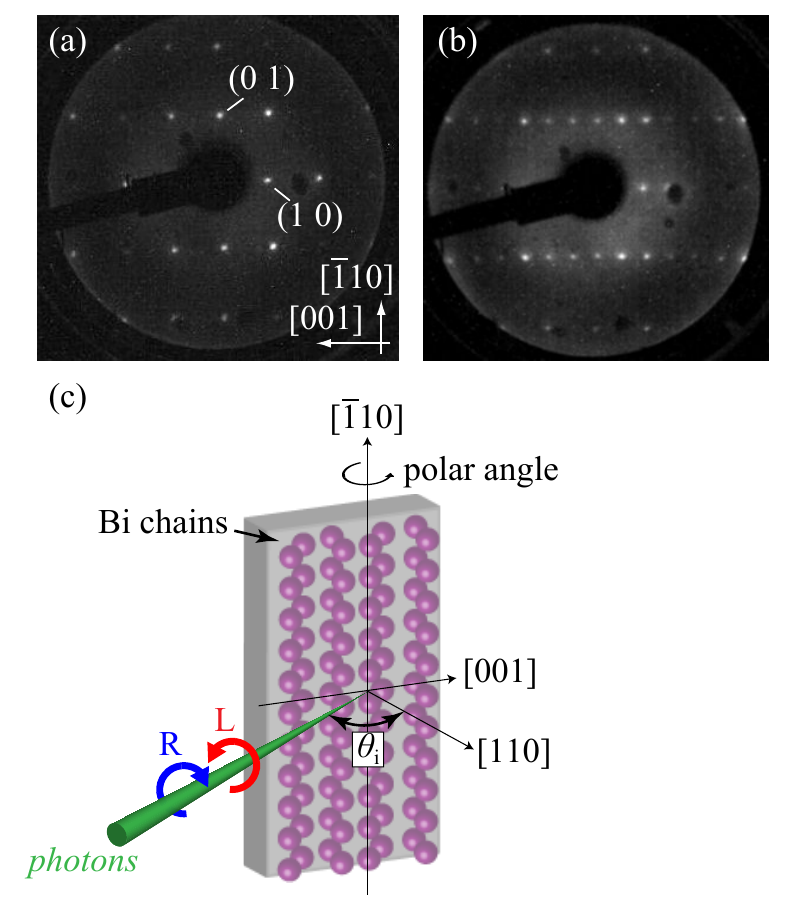}
\caption{\label{figure 1}
LEED patterns of the (a) clean GaSb(110) and (b) Bi/GaSb(110)-(2$\times$1) surfaces taken at room temperature.
Electron kinetic energy has been set as 75 eV for both patterns.
(c) A schematic drawing of the geometry of ARPES experimental setup together with the surface Brillouin zone of Bi/GaSb(110)-(2$\times$1) (see Section II for detailed explanation).
}
\end{figure}

%ARPES, VB
\begin{figure}[p]
\includegraphics[width=80mm]{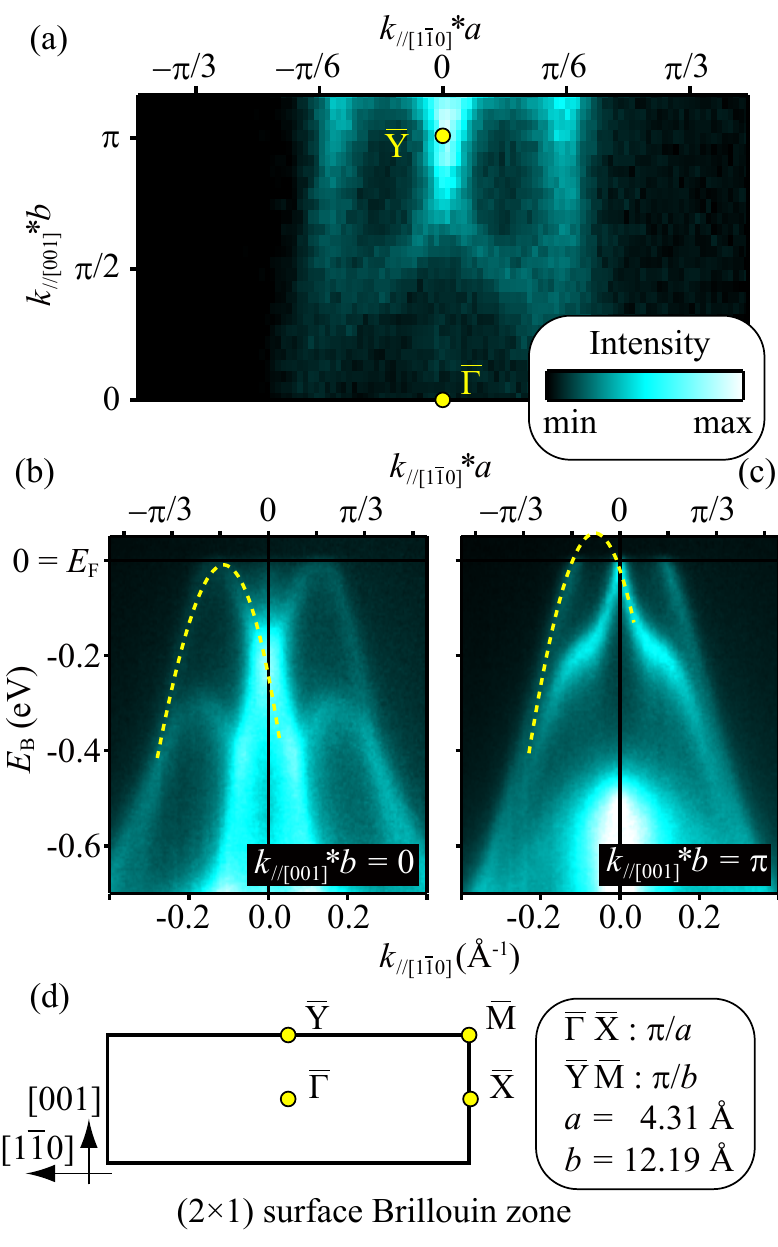}
\caption{\label{figure 2}
(a)--(c) ARPES intensity plots of Bi/GaSb(110)-(2$\times$1) at 25 K taken with circularly polarized photons ($h\nu$ = 30 eV).
The intensities from the left- and right-handed polarizations are summed up to show all the states without any influence from circular dichroism.
(a) Fermi contour, (b) valence band dispersions parallel to [1$\bar{1}$0] along $\bar{\Gamma}$--$\bar{\rm X}$, and (c) the same as (b) but along $\bar{\rm Y}$--$\bar{\rm M}$.
(d) (2$\times$1) SBZ and relevant surface lattice constants.
Dashed curves in (b) and (c) are the surface band dispersions simulated in Section III. D.
}
\end{figure}

%ARPES, CB
\begin{figure}[p]
\includegraphics[width=80mm]{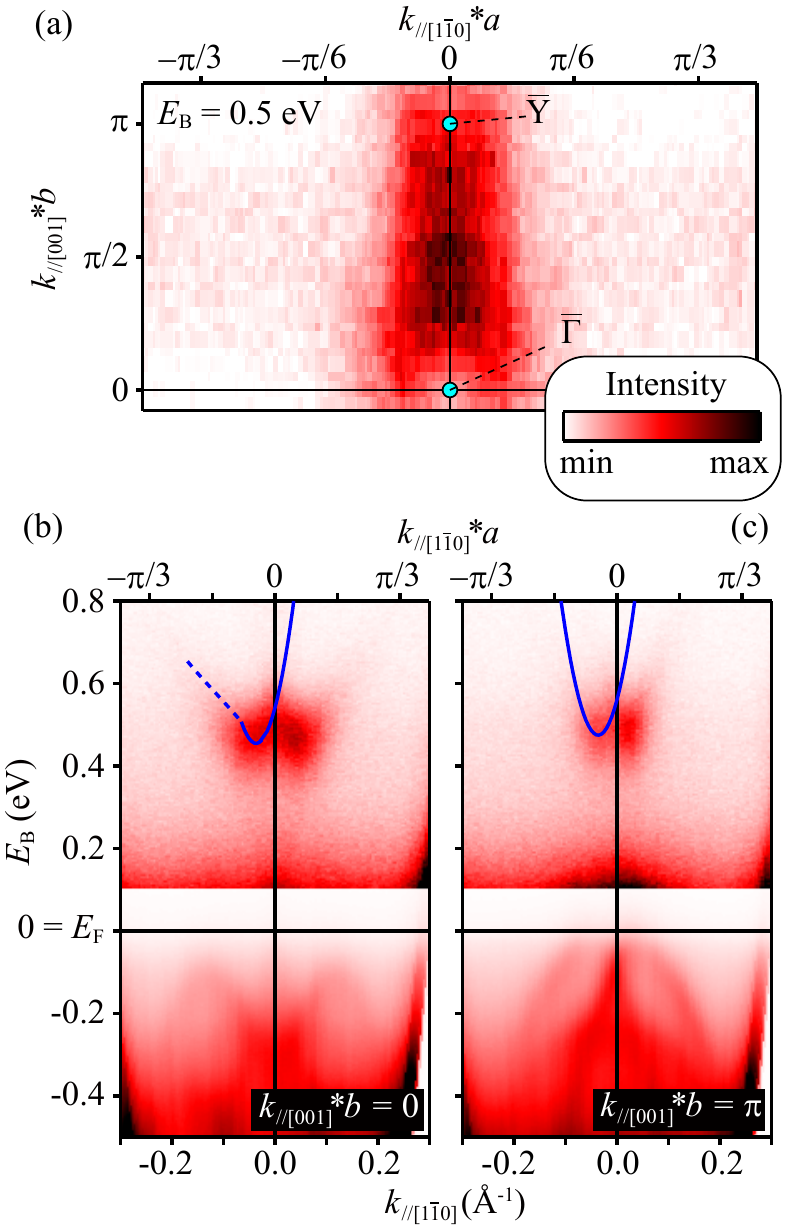}
\caption{\label{figure 3}
Same as Fig. 2 taken at 25 K but with two laser pulses. A delay time between the pump and probe pulses was set to 1 ps.
The pump pulses ($h\nu$ = 1.5 eV) were linearly polarized photons, the electric field vector, which lies in the photon incident plane.
The probe ones had the same incident plane as the circular polarization ($h\nu$ = 6.0 eV). The circular dichroism is canceled out in this figure as Fig. 2 and it is analyzed in Fig. 4.
(a) Constant energy contour at $E_{\rm B}\sim$ 0.5 eV.
(b) ARPES intensity plot along $\bar{\Gamma}$-$\bar{\rm X}$. The photoelectron signals in $E_{\rm B} >$ 0.1 eV are enhanced (60 and 80 times for (a) and (b), respectively) to make the surface conduction band dispersion visible.
(c) Same as (b) but taken along $\bar{\rm Y}$--$\bar{\rm M}$.
Solid and dashed curves in (b) and (c) guide the band dispersions. Solid curves are based on the theoretical calculation in Section III. D.
}
\end{figure}
%\begin{figure*}[p]

%CD
\begin{figure*}[p]
\includegraphics[width=150mm]{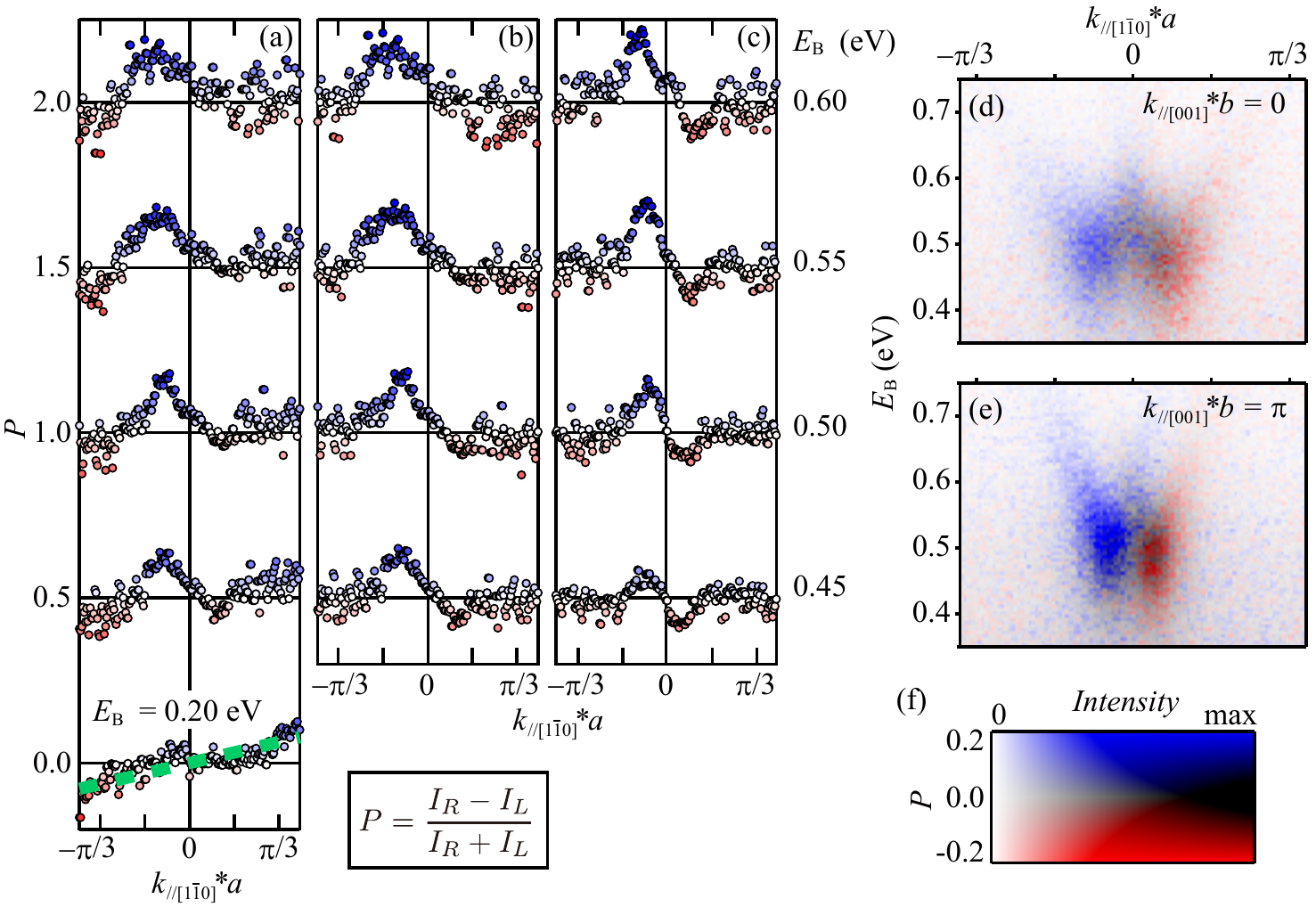}
\caption{\label{figure 4}
(a) Circular dichroism signals $P$ defined in the figure, taken along $\bar{\Gamma}$--$\bar{\rm X}$ at each $E_{\rm B}$, obtained from the same data as those shown in Fig. 3 (b).
$I_R$ ($I_L$) is the photoelectron intensity obtained with the right- (left-) handed probe photon polarizations, normalized by the ARPES intensity integrated along both $k_{[1\bar{1}0]}$ and $E_{\rm B}$ in the 2D region shown in Fig. 3(b).
The dashed line overlaid on the data for $E_{\rm B}$ = 0.20 eV is the estimated linear background (see the main text III. B for details).
Each $P$ curve is shown with an offset of 0.5.
(b) Same as (a) but the linear background is subtracted.
(c) Same as (b) but taken along $\bar{\rm Y}$-$\bar{\rm M}$.
(d), (e) Circular-dichroism ARPES plots, the color of which is determined by $P$ and the photoelectron intensities based on the color panel (f), taken along (d) $\bar{\Gamma}$--$\bar{\rm X}$ and (e) $\bar{\rm Y}$--$\bar{\rm M}$.
}
\end{figure*}

%CD
\begin{figure*}[p]
\includegraphics[width=150mm]{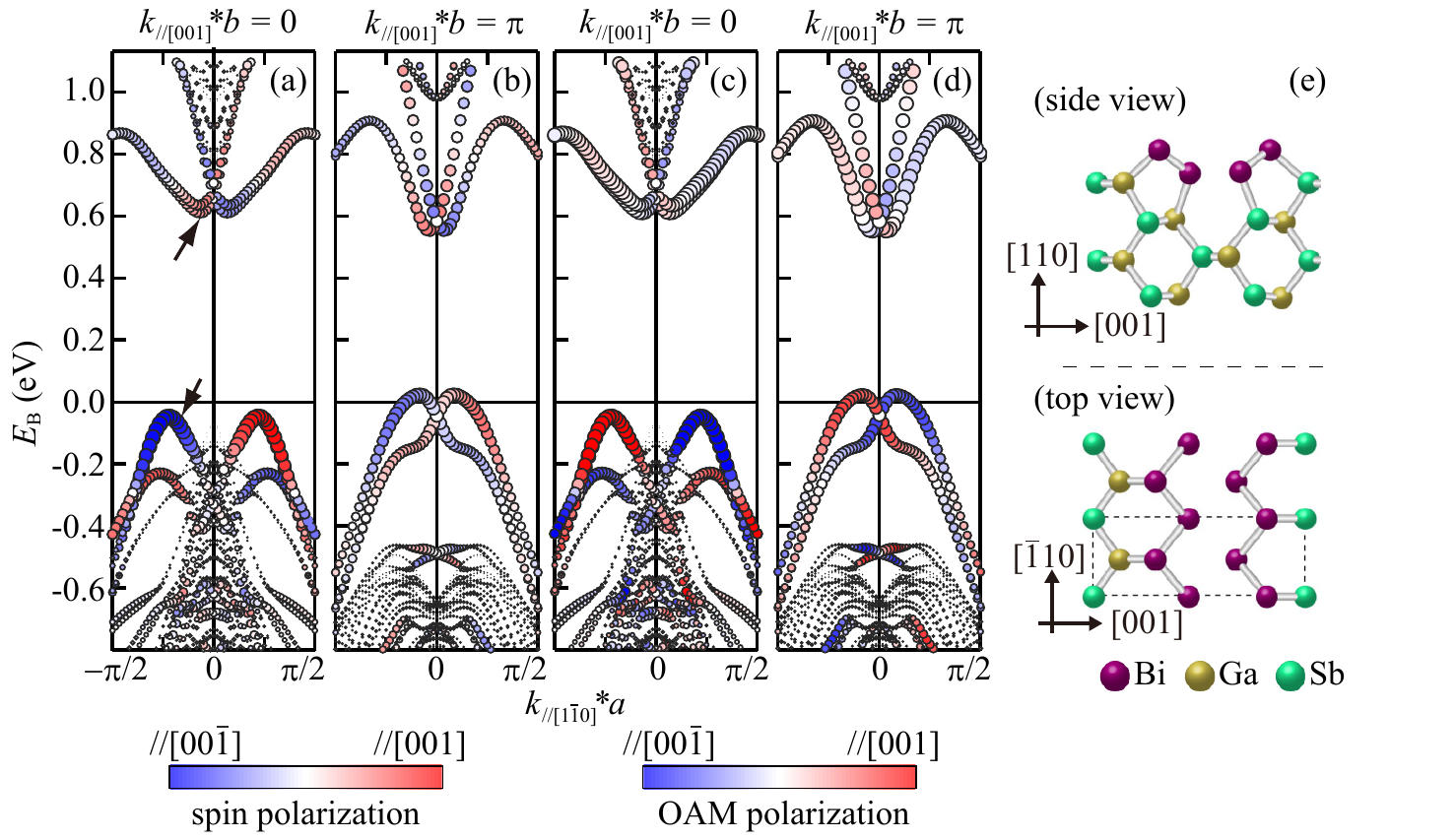}
\caption{\label{figure 5}
(a)-(d) Calculated band structures of Bi/GaSb(110)-(2$\times$1) along (a, c) $\bar{\Gamma}$-$\bar{\rm X}$ and (b, d) $\bar{\rm Y}$-$\bar{\rm M}$.
The radii of the circles are proportional to the contribution from the atomic orbitals of the surface Bi.
The contrasts (colors) of each circle in (a) and (b) show the spin polarization along [001], in-plane orientation perpendicular to $k_{[1\bar{1}0]}$.
Those in (c) and (d) are the OAM polarizations along [001].
(e) Surface atomic structure of Bi/GaSb(110)-(2$\times$1) used for the calculation. This figure is drawn from the same data as those shown in Fig. 1 (a) of ref. \onlinecite{Nakamura19}.
}
\end{figure*}

%ARPES, CB
\begin{figure}[p]
\includegraphics[width=80mm]{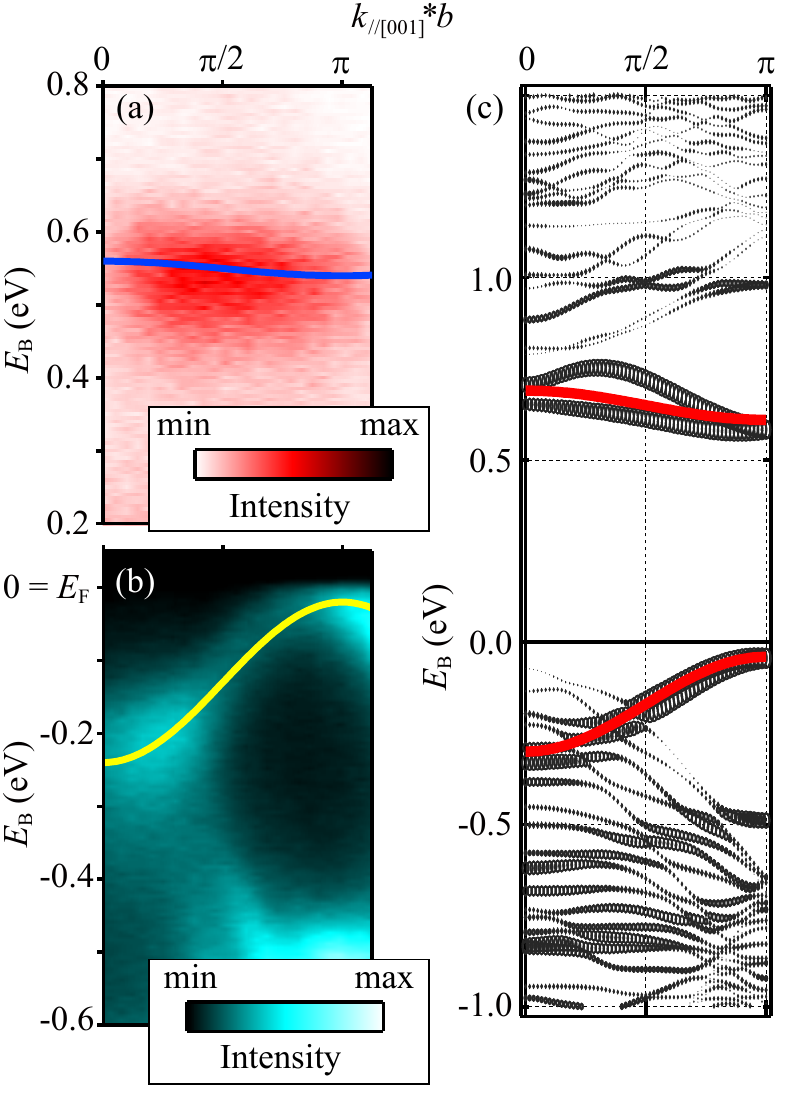}
\caption{\label{figure 6}
(a, b) ARPES intensity plots of Bi/GaSb(110)-(2$\times$1) along $\bar{\Gamma}$-$\bar{\rm Y}$ ($k_{[1\bar{1}0]}$*$a$ = 0) at 25 K. Solid curves are from the tight-binding model calculation (see Section III D in the main text).
(a) surface CB dispersion with two photon pulses (1.5 eV pump and 6.0 eV probe) at the delay time of 1.0 ps, and
(b) Surface VB dispersion at 25 K taken with one-photon ARPES (circularly polarized, $h\nu$ = 30 eV).
(c) Calculated band structure by the DFT calculation (the same condition as Fig. 5). The radii of the ellipses represents the contribution from surface Bi atoms. Solid curves are the surface CB and VB from the tight-binding model calculations.
}
\end{figure}
%\begin{figure*}[p]

% TB parameters
\begin{table}[p]
\caption{
Parameters for the tight-binding model calculations (eq. (1)) in units of eV.
}
\label{table2}
\begin{ruledtabular}
\begin{tabular}{lccc}
 & $t_a$ & $t_b$ & $\mu$ \\
\hline
ARPES, VB & 1.050 & 0.055 & -2.230 \\
ARPES, CB & -3.500 & -0.005 & 7.550 \\
DFT, VB & 1.050 & 0.065 & -1.930 \\
DFT, CB & -3.500 & -0.020 & 6.350
\end{tabular}
\end{ruledtabular}
\end{table}

% orbital fractions
\begin{table}[p]
\caption{
Calculated fractional contributions of atomic orbitals of surface Bi to the spin-split surface states $S_1$ for the conduction bands and $S_2$ for the valence bands (indicated by the arrows in Fig. 5(a)).
The $p_a$, $p_b$, and $p_c$ orbitals correspond to the $p$ orbitals, the lobes of which are along [1$\bar{1}$0], [001], and [110], respectively.
The values are normalized by the maximum 6$p_c$ for the valence band $S_2$.
}
\label{table2}
\begin{ruledtabular}
\begin{tabular}{lcccc}
% & \multicolumn{4}{c}{Distance (\AA)} \\
 & 6$s$ & 6$p_a$ & 6$p_b$ & 6$p_c$ \\
\hline
$S_1$ & 0.37 & 0.44 & 0.15 & 0.68 \\
$S_2$ & 0.24 & 0.41 & 0.63 & 1 
\end{tabular}
\end{ruledtabular}
\end{table}

\end{document}